# Using the Energy and Momentum Conceptual Survey to investigate progression in student understanding from introductory to advanced levels

Mary Jane Brundage[1], Alexandru Maries[2] and Chandralekha Singh[1]

[1] Department of Physics and Astronomy, University of Pittsburgh, Pittsburgh, PA, 15260

[2] Department of Physics, University of Cincinnati, Cincinnati, OH, 45221

**Abstract.** The Energy and Momentum Conceptual Survey (EMCS) is a multiple-choice survey that contains conceptual problems involving a variety of energy and momentum concepts covered in a typical introductory physics course for science and engineering majors. Prior studies suggest that many concepts on the survey are challenging for introductory physics students and the average student scores after traditional lecture-based instruction are low. The research presented here investigates the progression in student understanding of the EMCS concepts including their evolution from the beginning to the end of their courses in introductory and advanced level undergraduate physics after traditional lecture-based instruction. We find that on all EMCS questions on which less than 50% of the introductory physics students answered a question correctly after traditional instruction, less than two-thirds of the upper-level undergraduate students provided the correct response after traditional lecture-based instruction. We discuss the EMCS questions that remain challenging and the common alternate conceptions among upper-level students. The findings presented here are consistent with prior research showing that traditional instruction in upper-level courses, which typically focuses primarily on quantitative problem solving and often incentivizes use of algorithmic approaches, is not effective for helping many students develop a functional understanding of underlying concepts.

## INTRODUCTION

One goal of an introductory mechanics course for physical science and engineering majors is for students to develop a robust conceptual understanding of the underlying physics. For example, in the context of energy and momentum concepts, the University of Washington physics education group has conducted many investigations and developed research-based tutorials to improve student understanding [1-5], e.g., of work-kinetic energy and impulse-momentum theorems as well as the concept of systems in the context of energy and student ability to apply the concept of work and energy to extended systems. Van Heuvelen and Zou have emphasized the importance of multiple representations of work-energy processes, especially work-energy bar charts as a physical representation in problem solving to help students develop a good functional understanding of these concepts [6-8]. In addition to this type of focus on research and curriculum development, Seeley et al. [9-11] have examined physics teacher understanding of systems and the role it plays in supporting student energy reasoning and emphasized that physics teachers should not use the language suggesting that mechanical energy is a "conserved" quantity since it is the total energy that is a conserved quantity. Chabay et al. have proposed a unified, contemporary approach to teaching energy in introductory physics [12] and they have also suggested using mechanical energy is "constant" instead of "conserved" and emphasized that it is the energy that is a conserved quantity similar to Sealey et al. [9-11]. Ding et al. have designed an energy assessment to evaluate student understanding of energy [13] and found that students in an innovative principle-based mechanics course discussed in Ref. [12] showed significant improvement on the test objectives and were also able to reason conceptually on questions about energy without using equation sheets [14]. Moreover, since the presence of friction makes issues related to work and energy more challenging, Sherwood and Bernard have clarified issues related to work and heat transfer in the presence of sliding friction [15] and Daane et al. [16] have investigated teacher expectations and strategies associated with energy conservation in dissipative processes. Furthermore, since energy and momentum are important concepts in introductory physics, other investigators, e.g., Lin et al. used multi-part problems involving energy and momentum concepts to improve introductory physics students' problem-solving strategies [17, 18] and Yerushalmi et al. used similar problems to make the problem-solving process a learning opportunity by guiding students to diagnose their own mistakes in problem-solving [19, 20]. Also, since learning is context dependent, Barniol and Zavala have investigated the effects of different contexts on students' understanding of force, velocity and work using isomorphic problems [21] and Sagear et al. and Xu et al. have used research-based assessments involving different contexts to investigate student learning of energy and momentum concepts [22, 23].

Improving student conceptual understanding and problem solving remain major goals in courses for physics majors as they take advanced physics courses even though these courses use more complex mathematics to solve problems [24-28]. In the context of energy and momentum, students often learn about the basic concepts in the calculus-based introductory courses. Traditional upper-level undergraduate classical mechanics courses present these same concepts in the context of more complex, quantitative problems. Upper-level students are often expected to make connections between mathematical formalism and physics concepts themselves, without much support or incentive from instructors to integrate conceptual and quantitative aspects of the material [29, 30]. However, these complex quantitative problems can often be solved algorithmically without any course assessment specifically focusing on conceptual understanding. If advanced students in these traditionally taught courses are not given grade incentives and provided support to make the appropriate math-physics connections, they may solve problems by pattern matching. This skips the necessary step of unpacking the underlying concepts for developing a deep conceptual understanding of the foundational concepts, which is important for cultivating expertise in physics [29-33].

Several conceptual multiple-choice assessments have been developed and validated to evaluate student understanding of basic concepts covered in a typical introductory physics course [34-42]. Past research using other surveys has shown that even in upper-level undergraduate classes, students have difficulty with many introductory and advanced level concepts [43-45]. One such survey that we focus on here is the Energy and Momentum Conceptual Survey or EMCS [37], which has been used to investigate the conceptual understanding of students in introductory courses before and after traditional instruction in relevant concepts. It is a 25-question multiple-choice survey focused on energy and linear momentum. Here we discuss an investigation involving using the EMCS not only to evaluate the growth from the beginning (pre) to the end (post) in introductory and advanced mechanics courses but the evolution in student understanding of energy and momentum concepts from the introductory level to advanced level after instruction. We note that the EMCS uses the language typically used in the introductory physics courses and textbooks, e.g., that the mechanical energy is conserved when there is no work done on a system by non-conservative forces even though it has been emphasized that only the total energy is conserved. One should use the language that the mechanical energy of the system is constant instead of conserved and also refrain from using the term "non-conservative forces" since all forces are conservative at a fundamental level [9-12].

We also note that prior studies have sometimes identified issues including bias on questions for some conceptual assessments developed by physics education researchers. For example, on the Force Concept Inventory questions [34], McCullough found that gender gaps in performance can be influenced by switching the scenarios of the questions from stereotypically male to stereotypically female, although for some questions, the switch favored men, for others women, and yet for others, there was no difference [46, 47]. Several other investigations have found that specific questions on the FCI and Conceptual Survey of Electricity and Magnetism may be biased against women or men [48-51]. We are unaware of similar research using the EMCS.

Thus, limitations of this study are that the language used in the EMCS survey is consistent with traditional textbooks as noted earlier (e.g., mechanical energy of a system is conserved in certain situations) and we do not focus on bias, e.g., gender gap favoring men or women on specific EMCS questions. Instead, in this study, we empirically document the average performance on the EMCS before and after instruction in traditionally taught (primary lecture-based) introductory and advanced mechanics courses and analyze the evolution in student understanding of these concepts from introductory to advanced levels of all students enrolled in those courses.

## MOTIVATION AND GOAL FOR THIS STUDY

In this study, we empirically document the growth from the beginning to the end of calculus-based introductory and advanced undergraduate mechanics courses and analyze the progression in student conceptual understanding of energy and momentum using the EMCS. A typical goal of teaching mechanics at any level is to help advanced students understand the conceptual foundations at different levels of mathematical and conceptual sophistication. This helps students develop a better functional understanding of the underlying concepts while also acquiring mathematical facility in problem-solving. Conceptual validated surveys such as the EMCS can be useful tools for investigating the extent to which traditionally taught courses are succeeding in improving student conceptual understanding in introductory and advanced physics courses and the extent to which the performance of students in advanced undergraduate mechanics course is better than students in introductory physics. Using conceptual surveys in traditionally taught courses can help benchmark this evolution to compare it with future studies involving physics courses with research-based curricula in which there is an explicit focus on integrating conceptual and quantitative aspects of learning relevant concepts and helping students develop functional understanding. While conceptual assessments have been used heavily in physics education research, few studies have used them in the manner we do here. This study is useful not only for instruction in upper-level classical mechanics courses but also for introductory courses because knowledge about the concepts that are challenging for both student populations can help instructors at both levels contemplate why those concepts continue to be challenging and develop effective approaches for improving student conceptual understanding of classical mechanics throughout. We note that this study is not a longitudinal study in which one follows the same students but a cross-sectional study in which we analyze the evolution on EMCS concepts from the calculus-based introductory course to advanced mechanics course for the entire student population in those courses. In particular, our goal here is to evaluate changes from the beginning to the end of the courses and compare the overall effectiveness of the traditionally taught calculus-based introductory and advanced mechanics courses in helping students develop a conceptual understanding of the content covered in the EMCS.

**RESEARCH QUESTIONS**

To investigate changes from the beginning (pre) to the end (post) of the traditionally taught introductory and advanced undergraduate mechanics courses and the evolution in students' understanding and reasoning related to the EMCS, we investigated the following research questions:

RQ1. On which EMCS questions do the upper-level undergraduate students show moderate improvement or do very well in post-test compared to the pre-test?
RQ2. On which EMCS questions do upper-level undergraduate students struggle after traditional instruction, where "struggle" means that less than two-thirds of the students answered the question correctly? Are there any patterns in student responses from introductory courses to upper-level courses for these questions?
RQ3. What are the common alternate conceptions of upper-level students that cause them to struggle on the EMCS questions? Are there examples of persistent alternate conceptions from the introductory to the upper level?

**METHODOLOGY**

Both introductory and upper-level students involved in this study were from the same large public research university in the US. The data presented in all of the tables presented here were collected before the COVID-19 pandemic started. Both introductory and advanced mechanics courses were traditionally taught (primarily lecture based) with little effort to incorporate research-based curricula and pedagogies. The introductory students in this study were enrolled in a calculus-based physics course during their first year in college and were primarily engineering majors (~70%) with the rest being physics, chemistry, and mathematics majors. The introductory course used the textbook by Halliday, Resnick and Walker. The upper-level undergraduate classical mechanics course, which is required for physics majors, is usually taken by students in their third year. This course used Taylor's "Classical Mechanics" as the textbook. Data for this course were collected over two different consecutive years and averaged together to increase the statistical power since there were no contextual changes in the course across the two years. These students took the pre-test within the first week of classes and the post-test in the last few weeks of classes. All students used paper scantrons to bubble their responses to each question and were given approximately 50 minutes, which was sufficient time for everyone to answer all the questions. As noted earlier, this study is a cross-sectional study in which we analyze the evolution on EMCS concepts from the calculus-based introductory course to advanced mechanics course for the entire student population in those courses because our goal is to evaluate the overall effectiveness of the traditionally taught calculus-based introductory and advanced mechanics courses in helping students develop a conceptual understanding of the content covered in the EMCS.

Both groups of students took the EMCS as a pre-test (before instruction in relevant concepts) and as a post-test (after instruction). In the introductory physics course [52], 352 students took the pre-test and 336 students took the post-test. In the upper-level undergraduate course, 68 students took the written pre-test and 42 students took the post-test. We investigated changes from the beginning (pre) to the end (post) of the traditionally taught introductory and advanced undergraduate mechanics courses using normalized gains on each question, i.e., (post%-pre%)/(100%-pre%) [53]. In particular, we compared the average normalized gains of advanced and introductory students. Then, we specifically focused on the performance of upper-level students on the EMCS at the end of instruction with their performance at the beginning to identify questions on which they showed moderate improvement. We then compared in detail the post-test performance of the upper-level undergraduates with the post-test performance of introductory physics students. We focused on identifying concepts which upper-level students still had difficulties with at the end of a traditionally taught upper-level classical mechanics course and analyzed different patterns of evolution across the EMCS questions from the introductory to the upper-level, e.g., the type of questions on which advanced students improved significantly or did not do so. In particular, we investigated the EMCS concepts that were challenging for both introductory and upper-level students and the EMCS concepts that were not mastered at the introductory level but were mastered at the upper-level. Additionally, we analyzed data to investigate if there are EMCS questions on which the alternate conceptions of introductory students are different from the alternate conceptions of upper-level students or if a dominant alternate conception is observed for advanced students (as opposed to being spread across multiple alternate conceptions among introductory students). For the questions on which the upper-level students struggled, we identified potential reasons for why many students were unable to answer those questions correctly in those contexts after traditional instruction by asking a different set of 26 upper-level undergraduate students in a mechanics course at the same university to explain their reasoning for the answers they provided to each of the EMCS questions. The quotations provided from these students in the results section are the most common responses except in cases in which it was difficult to understand student explanations or there were only a few students in that particular category.

Although researchers can use different criteria, it is useful to adopt criteria, e.g., for the average EMCS scores or normalized gains that can be considered to be fair performance or good performance. We adopted the following criteria:

- If at least two-thirds of the students in a group provide the correct response to a particular EMCS question, their overall performance on that question would be considered fair. This criterion is similar to Hestenes' criterion of 60% score to be considered a threshold for Newtonian understanding on surveys he developed, e.g., see [54].
- If 80% or more of the students answer an EMCS question correctly, it would correspond to good performance. This cutoff was chosen in part because for any multiple-choice test, it is possible for students to not read the information carefully or make mistakes due to simply being inattentive and not due to not understanding the concept.
- If the average normalized gain is 0.3 or larger on a particular question, this is considered a fair improvement from the beginning to the end of a course. This is between the overall FCI normalized gains [53] for traditionally taught ($g=0.23$) and interactive engagement courses ($g=0.48$).

We present data from all upper-level students who took the written pre/post-tests because analysis using only matched data for pre/post-tests yields similar outcomes without practically significant differences. For example, for the upper-level students, the pre-test averages for matched and unmatched data are 63.3% and 63.1% and the post-test averages for matched and unmatched data are 68.4% and 67.4%, respectively.

**RESULTS**

We note that all the EMCS questions are provided in the appendix of Ref. [37]. Below, we answer each of the research questions:

***RQ1. On which EMCS questions do the upper-level undergraduate students show moderate improvement or do very well in post-test compared to the pre-test?***

Table I shows the pre-test and post-test performance of introductory and upper-level students for each question on the EMCS and Fig. 1 shows the post-test scores for both levels for each of the 25 questions. To answer RQ1, we used two measures. One measure that we used to quantify the improvement from the pre-test to post-test (see Table I) is normalized gain [53], which is a common measure used to quantify performance changes relative to pre-test performance. We focused on questions in which the average normalized gain (called normalized gain for short from now on) for the improvement from the pre-test to the post-test was 0.3 or larger, i.e., students improved by 30% relative to their pre-test performance.

Table I shows that upper-level students had normalized gains of 0.3 or larger on 7 EMCS questions while introductory students had normalized gains in that range on 11 out of 25 questions. This may be because, as shown in Table I, the average pre-test scores were lower for some of the questions for the introductory students. This distinction can also be seen in Fig. 2 and Fig. 3, in which we plot the pre-test and post-test scores on the vertical axis and the normalized gain on the horizontal axis for the upper-level and the introductory level students, respectively. The cutoff of 0.3 for normalized gain was also in part chosen because there appears to be a natural cutoff in the upper-level student data around 0.3. In particular, Table I shows that when ranking the questions according to the normalized gains for upper-level students, we find seven questions with a decrease in performance (so normalized gain is not recommended to be used), then normalized gains of 0.07, 0.10, 0.11, 0.15, 0.22, 0.23, 0.24, 0.24, 0.26, 0.28, and 0.29.

As soon as we pass 0.3, we have 0.32, 0.35, and all the rest have normalized gains of 0.4 or more.

In addition to using the normalized gain, we used another measure to answer RQ1. As noted, we adopted a standard that 80% or more of the students answering an EMCS question correctly would correspond to good performance. Also, in our data, there appears to be a natural cutoff around 80% for the upper-level students: ranking from highest performance to lowest performance, we have 95%, 90%, 90%, then six questions with performance between 80% and 90%, after which two questions with performance of 76% and everything else is below 70%. We note that on nearly all questions in which upper-level students had good performance on the post-test, their pre-test performance was also reasonably high. For comparison, Table I shows that introductory students only exhibited performance greater than 80% on two questions and there was one question on which their post-test performance was 79%.

**Table I.** Pre- and post-test performance of introductory and upper-level undergraduate students (in percent) for each question on the EMCS along with normalized gains (g) for the change from the pre-test to the post-test. We note that on the questions in which student performance deteriorates, the normalized gain is not calculated, a common practice for using this measure [53]. Thirteen questions on which the post-test performance of upper-level students is low (below two-thirds correct) are labeled with (L) next to the item number, nine questions on which the post-test performance is high (above 80% correct) are labeled with (H) next to the item number, and three questions with medium performance, between 71%-76%, are given a label (M) next to the item number. The standard errors for the correct responses for the introductory students range from 2% to 3% and the upper-level students range from 3% to 8% for both pre- and post-tests across all 25 questions.

| EMCS Q# | | Introductory Students | | | Upper-Level Students | | |
|---|---|---|---|---|---|---|---|
| | | Pre (N=352) | Post (N=336) | Gain | Pre (N=68) | Post (N=42) | Gain |
| 1 | (H) | 33% | 63% | 0.45 | 76% | 88% | 0.49 |
| 2 | (H) | 50% | 81% | 0.63 | 76% | 83% | 0.29 |
| 3 | (L) | 45% | 61% | 0.30 | 65% | 48% | … |
| 4 | (H) | 47% | 69% | 0.40 | 71% | 86% | 0.51 |
| 5 | (L) | 11% | 27% | 0.18 | 47% | 52% | 0.10 |
| 6 | (L) | 12% | 28% | 0.18 | 47% | 38% | … |
| 7 | (H) | 55% | 85% | 0.66 | 87% | 90% | 0.28 |
| 8 | (L) | 30% | 45% | 0.22 | 57% | 62% | 0.11 |
| 9 | (L) | 26% | 36% | 0.13 | 49% | 62% | 0.26 |
| 10 | (L) | 13% | 35% | 0.25 | 57% | 55% | … |
| 11 | (H) | 41% | 79% | 0.65 | 85% | 95% | 0.68 |
| 12 | (L) | 21% | 33% | 0.15 | 47% | 60% | 0.24 |
| 13 | (L) | 52% | 53% | 0.03 | 60% | 57% | … |
| 14 | (H) | 60% | 70% | 0.25 | 90% | 83% | … |
| 15 | (M) | 47% | 58% | 0.22 | 69% | 76% | 0.23 |
| 16 | (L) | 34% | 26% | … | 44% | 62% | 0.32 |
| 17 | (L) | 30% | 46% | 0.23 | 54% | 64% | 0.22 |
| 18 | (H) | 20% | 66% | 0.57 | 82% | 90% | 0.46 |
| 19 | (M) | 47% | 70% | 0.43 | 72% | 71% | … |
| 20 | (H) | 35% | 58% | 0.34 | 78% | 81% | 0.15 |
| 21 | (H) | 44% | 62% | 0.32 | 71% | 83% | 0.43 |
| 22 | (L) | 16% | 29% | 0.16 | 32% | 29% | … |
| 23 | (L) | 21% | 25% | 0.06 | 51% | 55% | 0.07 |
| 24 | (L) | 20% | 29% | 0.11 | 50% | 62% | 0.24 |
| 25 | (M) | 33% | 59% | 0.39 | 63% | 76% | 0.35 |

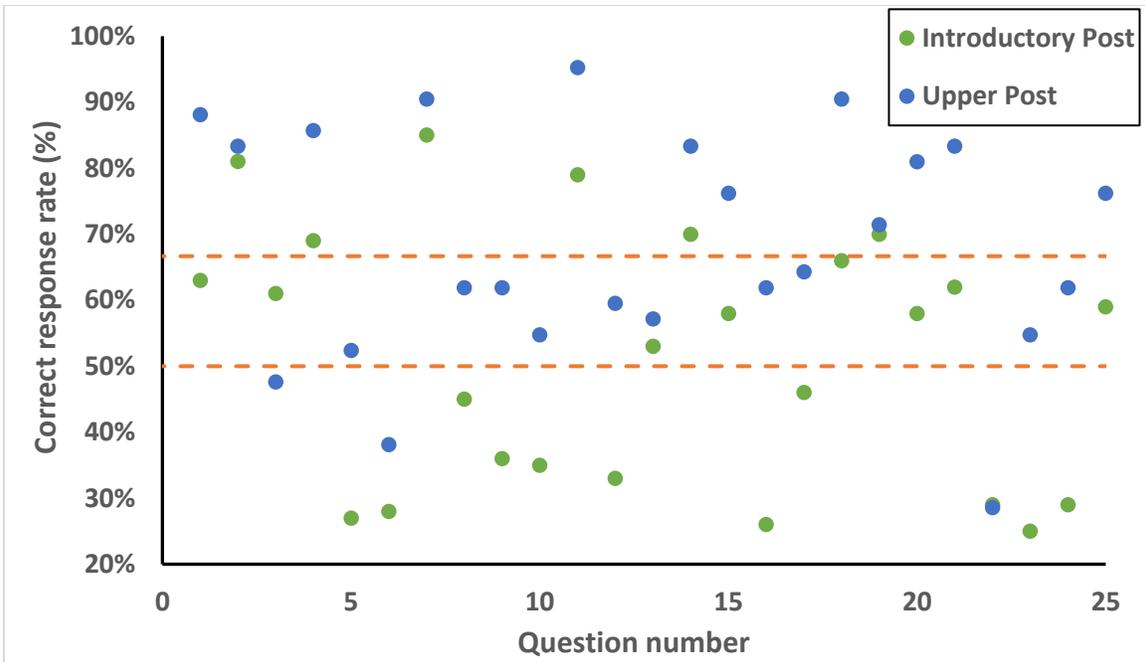

Fig. 1. Average post-test scores for upper-level and introductory-level students on each question. Data from the introductory-level students is shown in green while the upper-level data is in blue. Each question in which introductory level students had less than 50% correct, the upper-level students did not reach two-thirds correct. The 50% correct and two-thirds correct are shown with orange dashed lines.

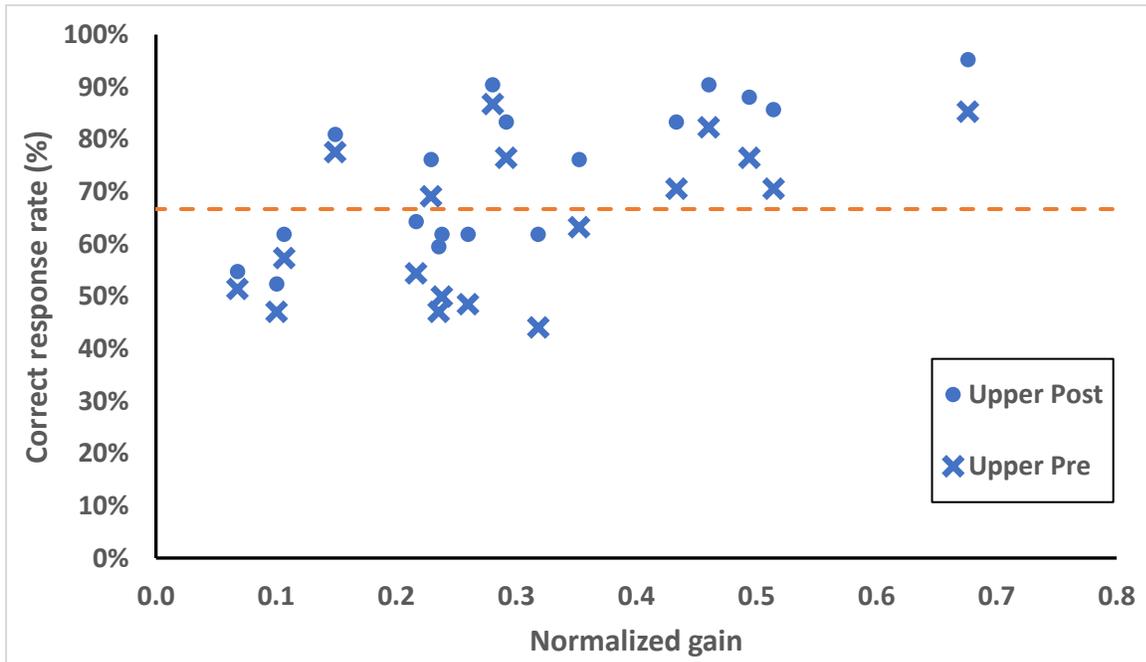

Fig. 2. Upper-level student scores vs normalized gain. Normalized gain is calculated using both the pre-test and post-test scores, so for each question, the pre-test and post-test scores appear vertically stacked on the graph. The post-test scores are shown with circles and the pre-test scores are shown with the Xs. The two-thirds correct criterion for fair performance for upper-level students is shown with an orange dashed line.

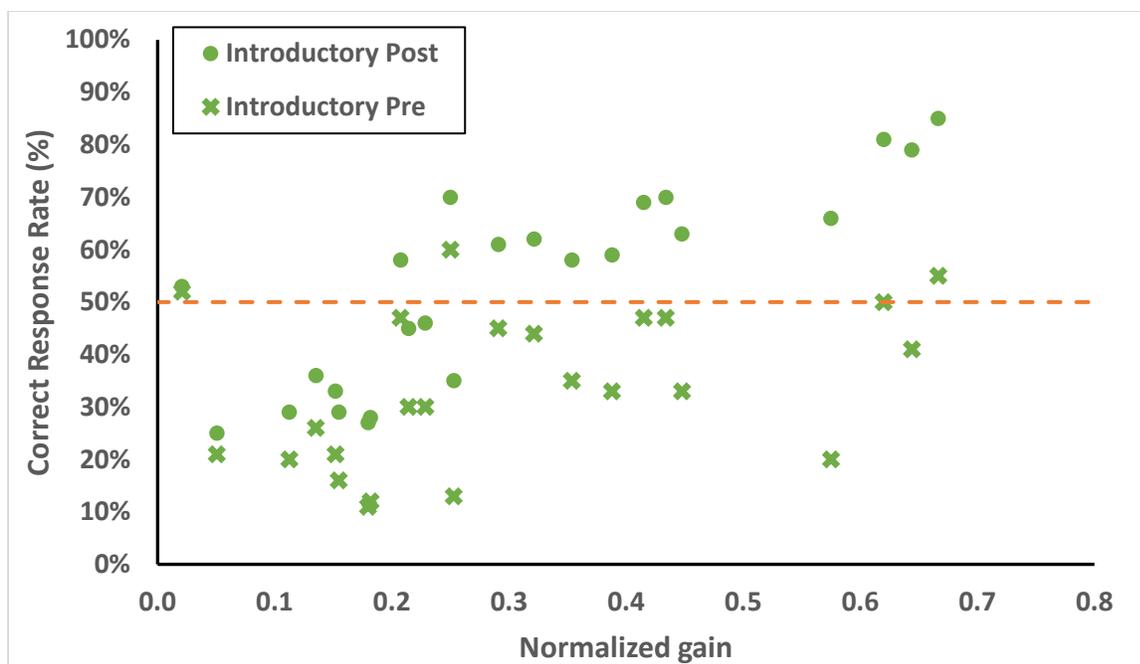

Fig 3. Introductory-level student scores vs normalized gain. Normalized gain is calculated using both the pre-test and post-test scores, so for each question, the pre-test and post-test scores appear vertically stacked on the graph. The post-test scores are shown with circles and the pre-test scores are shown with the Xs.

Both these measures (post-test performance of 80% or more and normalized gain of 0.3 or higher) were used because if the pre-test performance of upper-level students is high on a question, there isn't much room for improvement, and it is unreasonable to expect a large, normalized gain. The following questions fulfill at least one of these two criteria for upper-level students: Q1, Q2, Q4, Q7, Q11, Q14, Q16, Q18, Q20, Q21 and Q25. On nearly all of these questions, the pre-test performance of upper-level students was higher than 70%; the only exceptions are Q16 and Q25, on which, even though the upper-level students exhibited normalized gains larger than 0.3, their post-test performance was not higher than 80%. This suggests that the upper-level classical mechanics course did little to improve many students' conceptual understanding of energy and momentum as measured by their performance on these EMCS questions. This lack of significant improvement is also reflected in the fact that the overall normalized gain for the entire EMCS survey for upper-level students was 14% (from an overall average of 63% to 68%).

Q11 exhibits the largest normalized gain (0.68), but upper-level student improvement was from 85% to 95%, suggesting that the question was easy both in the pre-test and in the post-test. The question with the next largest normalized gain is Q4 (improvement from 71% to 86%). In Q4, two stones are thrown from the same height, one straight up and the other straight down, and students are asked which stone has the greater speed when reaching the ground. Many upper-level students recognized that to answer the question, they could either use the rule that mechanical energy is constant or that when the stone thrown upwards reaches the same height from where it was thrown, it is moving downwards at the same speed, so it is indistinguishable from the stone thrown downwards. For example, one student stated "Conservation of energy. Both balls have same initial kinetic/potential energy, so they should finish with the same speed if they fall the same height." Another student stated that "when [stone] A reaches its original position while falling, the velocity is the same as the initial velocity, but opposite direction, meaning it is the same as [stone] B." These were very typical responses on this question.

On Q1, the normalized gain for the upper-level students was 0.49, corresponding to an improvement from 76% to 88%. This indicates that on the post-test, students were more likely to recognize that the work done by the gravitational force only depends on the height difference and not on the path or how quickly the object is lifted. Among the students who provided explanations, they either motivated their answer by relating the work done by the gravitational force to the change in the gravitational potential energy or made note of the fact that the force of gravity is a conservative force, and so the work done by it is path independent. For example, one student stated that "work is change in potential energy and PE [potential energy] only depends on weight and height" and another student stated that "[the force of] gravity is a conservative force and therefore work done is path independent, so the only distance that matters is the distance travelled from the floor to the table."

Next is Q18 which had a normalized gain of 0.46, but the improvement was from 82% to 90%, meaning that this was a relatively easy question for the upper-level students both in the pre-test and the post-test. Next, Q21 had a normalized gain of 0.43 for upper-level students due to an improvement from 71% to 83% indicating that in the post-test, students were more likely to recognize that a cart moving horizontally into which rain falls directly downwards will slow down due to conservation of momentum. Nearly all the students who provided explanations mentioned that conservation of momentum implies that the speed of the cart decreases because its mass increases. For example, "Momentum must be conserved, so if mass of the cart goes up then the speed must go down" was a typical explanation given by the upper-level students.

The next highest normalized gain of 0.35 for upper-level students is for Q25, corresponding to an improvement from 63% to 76%. Q25 describes a situation in which a box moving with some initial speed on a horizontal surface slows down and stops and asks which of the following is equal to the change in kinetic energy of the box among the following choices: (a) the momentum of the box multiplied by the distance traveled before coming to rest, (b) the momentum of the box multiplied by the time elapsed before coming to rest, (c) the momentum of the box multiplied by the deceleration of the box, (d) the mass of the box multiplied by the deceleration of the box, and (e) none of the above. After instruction, upper-level students were more likely to recognize that none of the options are correct (none of the options have the same units as kinetic energy). When providing written explanations, the majority of students mentioned units, for example, one student stated, "none of the options match the dimensions of KE [kinetic energy]."

Q16 provides the diagram shown in Fig. 4. Sphere A is raised to a height $h_0$ and released; it collides with sphere B at the bottom and the two stick together and swing to a maximum height $h_f$. The question asks for the principles which are needed to solve the problem. On this question, the normalized gain of upper-level students was 0.32. This is also the question with the largest improvement for upper-level students from the pre-test to the post-test (and the only question with an improvement greater than 15% from 44% to 62%). We note that this was still a difficult question for the upper-level students in the post-test, and we discuss students' most common difficulties when answering RQ3.

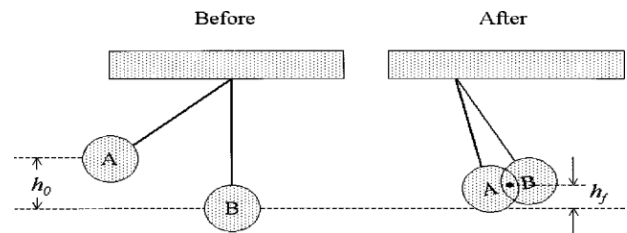

**Fig. 4.** Diagram for Q16.

*RQ 2. On which EMCS questions do upper-level undergraduate students struggle after traditional instruction, where "struggle" means that less than two-thirds of the students answered the question correctly? Are there any patterns in student responses from introductory courses to upper-level courses for these questions?*

To answer RQ2, we analyzed data to identify patterns in student responses across these questions with regard to the performance of both introductory and upper-level students. Table I shows that there are 13 questions on which the performance of the upper-level students is below two-thirds correct (in Table I: Q3, Q5, Q6, Q8, Q9, Q10, Q12, Q13, Q16, Q17, Q22, Q23, Q24). These questions are easily identifiable in Fig. 1 in which all of the questions in which the introductory level students did not reach 50% correct, the upper-level students did not reach two-thirds correct. This is roughly half of the questions on the EMCS and indicates that the upper-level students are far from having developed a robust conceptual understanding of energy and momentum concepts after traditional instruction in an upper-level classical mechanics course. Comparing the post-test performance of the upper-level students with the corresponding post-test performance of introductory students on these questions, we find that for nearly all of them (11 out of 13), introductory students' post-test performance after instruction was less than 50%. In fact, on **all** of the questions on which the performance of introductory students is below 50% on the post-test, the performance of the upper-level students after instruction is below two-thirds, indicating that the concepts covered in those questions are very challenging for even the upper-level students to grasp after traditional instruction.

*RQ3. What are the common alternate conceptions of upper-level students that cause them to struggle on the EMCS questions? Are there examples of persistent alternate conceptions from the introductory to the upper level?*

To answer RQ3, we identified the most common alternate conceptions of students in both groups by analyzing the percentage of students who selected each answer choice. Here, we discuss questions in which the upper-level students struggled, with a particular focus on questions in which a significant fraction of upper-level students exhibited a specific alternate conception. Additionally, to answer the second part of this research question, we discuss situations in which students at both levels displayed persistent difficulties or there was only a small difference between the alternate conceptions that are most common among one group compared to the other. We note that in most cases, the most common incorrect answer choices of upper-level students were the same as for the introductory students but with lower percentages. In Table II, we show the percentages of introductory students (post-test) and upper-level students (post-test) who selected each answer choice. Although in RQ3 we focus on post-tests, in the Appendix (Table III), we show similar data as in Table II but for the pre-test for introductory and upper-level students for comparison with post-test for those interested in changes in alternate conceptions from the beginning to the end of the course.

**Table II.** Percentages of introductory students and upper-level students who selected each answer choice on each question on the EMCS after instruction (post-test). The correct answers are bold, underlined, and italicized. The N in the parentheses is the number of students in each group. The standard errors for the correct responses for the introductory students range from 2% to 3% and the upper-level students range from 3% to 8% across all 25 questions.

| EMCS Q. No | Introductory Students (N = 336) | | | | | Upper-Level Students (N = 42) | | | | |
|---|---|---|---|---|---|---|---|---|---|---|
| | A | B | C | D | E | A | B | C | D | E |
| 1 | 4 | _**63**_ | 21 | 6 | 6 | 5 | _**88**_ | 5 | 2 | 0 |
| 2 | 7 | 2 | 6 | 4 | _**81**_ | 14 | 2 | 0 | 0 | _**83**_ |
| 3 | 3 | _**61**_ | 3 | 21 | 12 | 0 | _**48**_ | 5 | 36 | 12 |
| 4 | _**69**_ | 9 | 3 | 8 | 11 | _**86**_ | 2 | 0 | 2 | 10 |
| 5 | 20 | 30 | 12 | _**27**_ | 11 | 7 | 29 | 7 | _**52**_ | 5 |
| 6 | 3 | 8 | _**28**_ | 37 | 24 | 0 | 10 | _**38**_ | 31 | 21 |
| 7 | 5 | 4 | 2 | 4 | _**85**_ | 5 | 0 | 2 | 2 | _**90**_ |
| 8 | 6 | 6 | _**45**_ | 33 | 10 | 0 | 10 | _**62**_ | 29 | 0 |
| 9 | _**36**_ | 24 | 12 | 4 | 24 | _**62**_ | 14 | 7 | 0 | 17 |
| 10 | 14 | 14 | 18 | _**35**_ | 19 | 0 | 5 | 12 | _**55**_ | 29 |
| 11 | 3 | 6 | 6 | 6 | _**79**_ | 0 | 2 | 0 | 2 | _**95**_ |
| 12 | 2 | 2 | 12 | _**33**_ | 51 | 2 | 2 | 14 | _**60**_ | 21 |
| 13 | 7 | 14 | _**53**_ | 9 | 17 | 7 | 17 | _**57**_ | 12 | 7 |
| 14 | 5 | 7 | 3 | _**70**_ | 15 | 0 | 2 | 7 | _**83**_ | 7 |
| 15 | _**58**_ | 2 | 31 | 5 | 4 | _**76**_ | 2 | 19 | 2 | 0 |
| 16 | 37 | 19 | _**26**_ | 6 | 12 | 14 | 21 | _**62**_ | 2 | 0 |
| 17 | 27 | _**46**_ | 2 | 23 | 2 | 17 | _**64**_ | 0 | 19 | 0 |
| 18 | 9 | 5 | 15 | 5 | _**66**_ | 2 | 5 | 0 | 2 | _**90**_ |
| 19 | 19 | _**70**_ | 3 | 4 | 4 | 21 | _**71**_ | 5 | 2 | 0 |
| 20 | _**58**_ | 7 | 12 | 11 | 12 | _**81**_ | 2 | 2 | 7 | 7 |
| 21 | 10 | 5 | _**62**_ | 11 | 12 | 10 | 2 | _**83**_ | 2 | 2 |
| 22 | 39 | 5 | 16 | _**29**_ | 11 | 33 | 2 | 31 | _**29**_ | 5 |
| 23 | 50 | _**25**_ | 19 | 4 | 2 | 29 | _**55**_ | 14 | 0 | 2 |
| 24 | _**29**_ | 24 | 21 | 20 | 6 | _**62**_ | 14 | 17 | 7 | 0 |
| 25 | 8 | 9 | 12 | 12 | _**59**_ | 5 | 7 | 2 | 10 | _**76**_ |

Q3 describes a situation in which a white hockey puck collides elastically with a red hockey puck (and no net external forces act on the two-puck system) and asks which statements are correct regarding the collision among (1) the kinetic energy of the white puck is conserved, (2) the linear momentum of the white puck is conserved, and (3) the linear momentum of the two-puck system is conserved. On this question, 36% of upper-level students selected statements (1) and (3) – answer choice D. It is noteworthy that the percentage of introductory students who selected this answer choice was significantly smaller (21%). It is possible that some of the students with this type of response associated "elastic collision" with "kinetic energy being constant" and did not think carefully about the fact that it is the kinetic energy of the _system_ which is constant. For example, one upper-level student justified their choice by saying "No external forces so momentum of the system is conserved, and it is [an] elastic [collision] so kinetic energy is conserved", and another stated "In elastic collision kinetic energy remains conserved. [The] Linear momentum of system (both white and red puck) remains conserved." These students stated that the linear momentum *of the two-puck system* is conserved, but for kinetic energy, they simply made a general statement that kinetic energy is conserved in an elastic collision. All of the students who selected this incorrect answer choice gave one of two lines of reasoning. Some simply stated something similar to this student, "conservation of energy and momentum" while others provided explanations that are similar to the quotes above. These students explicitly pointed out that it is the linear momentum of the two-puck system that is conserved while at the same time stating that the kinetic energy is conserved.

Q3 is also intriguing because it's the only question on the EMCS with a significant *decrease* in performance from 65% on the pre-test to 48% on the post-test: a decrease of 17%! This is the only question which had a decrease from the pre-test to the post-test larger than 10%. Additionally, it is the only question in which the upper-level students had a lower performance than the introductory students (48% compared to 61%) and also exhibits the third lowest performance on the EMCS with only half of the upper-level students answering this question correctly after instruction.

Q5 relates to the impulse momentum theorem: two bullets are shot towards blocks of equal mass, one made of wood and the other made of steel. The bullet embeds in the wood block and bounces off of the steel (elastically), and students are asked which block travels faster after the collision. Only roughly half of the upper-level students answered this question correctly and 29% of students selected that the wood block would travel faster because the bullet transfers all of its kinetic energy to it. Based on upper-level student explanations for those who selected this answer choice, it appears that the most common reasonings are similar to this student who stated that "the wood block now has the KE of the bullet instead of splitting it between two objects," and another who stated that "all of the bullet's energy goes into the wood, while the bullet is still moving for the steel, that implies the velocity of the wood block will be more." However, when the bullet embeds in the block, some of its kinetic energy is used to deform the block and lost to heat, so the kinetic energy of the system is not constant. This type of reasoning suggests that some of the students may have difficulty recognizing that only the momentum of the system is conserved in a perfectly inelastic collision. It is also worthwhile pointing out that the percentages of introductory students and upper-level students who had this alternate conception are nearly identical (30% for introductory students compared to 29% for upper-level students). Additionally, analyzing students' choices, introductory students selected that the wooden block would be moving faster after the collision for two reasons: 1) because it gains the kinetic energy of the bullet (30%), and 2) because it gains the momentum of the bullet (30%). Upper-level students who expected that the wooden block would be moving faster primarily selected the first reason at nearly the same rate as introductory students (29%), and only 7% selected the second.

Q6, the second most challenging question on the EMCS, provides a diagram showing the circular orbit of a satellite around the Earth (see Fig. 5) and asks which statement is correct among the following: (a) the gravitational potential energy of the satellite decreases as it moves from A to B, (b) the work done on the satellite by the gravitational force is negative for the motion from A to B, (c) the work done on the satellite by the gravitational force is zero for the motion from A to B, (d) the velocity of the satellite remains unchanged as it moves from A to B, and (e) none of the above. In this context, the gravitational potential energy of the satellite refers to the gravitational potential energy of the satellite-earth system. Q6 is a very difficult question for upper-level students: only 38% answered it correctly, a performance only 10% higher than that of introductory students (28%). On this question, the most common difficulties of both upper-level and introductory students were the same: 31% of upper-level and 37% of introductory students selected (d) and 21% of upper-level and 24% of introductory students selected (e). Students who selected (d) typically did not distinguish between velocity and speed. For example, one upper-level student who selected this answer choice motivated their choice by saying "It must move with constant velocity to have a circular path".

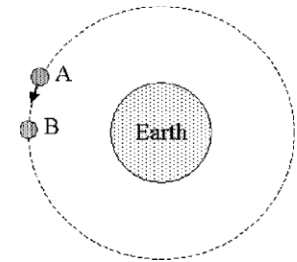

**Fig. 5.** Diagram for Q6.

Q8 describes the two situations shown in Fig. 6 in which a block is lifted at constant velocity over a height $h$, first directly up and second along a frictionless inclined plane. Students are asked to select the correct statement among: (a) the magnitude of the tension force is smaller in case (i) than in case (ii), (b) The magnitude of the tension force in the string is the same in both cases, (c) the work done on the block by the tension force is the same in both cases, (d) the work done on the block by the tension force is smaller in case (ii) than in case (i), (e) the work done on the block by the gravitational force is smaller in case (ii) than in case (i). On this question, 29% of upper-level undergraduate students selected (d), a percentage that is nearly identical to that of introductory students who selected the same answer (33%). Some students who selected that the work done by the tension force is less in the second case were likely focusing on the fact that the tension force required to lift the block at constant velocity along the incline plane is smaller in case (ii) than in case (i) and not consider that the distance over which the force acts is larger in case (ii) than in case (i). For example, one upper-level student noted that answer choice (d) is correct because "the magnitude of the tension force is smaller in case ii. Therefore, the work done is smaller in case ii."

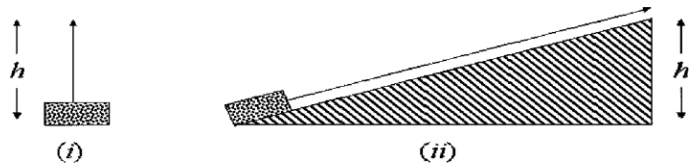

**Fig. 6.** Diagram for Q8.

Q9 describes three situations in which a bicyclist approaches a hill: (1) Cyclist 1 stops pedaling at the bottom of the hill, and her bicycle coasts up the hill, (2) Cyclist 2 pedals so that her bicycle goes up the hill at a constant speed, and (3) Cyclist 3 pedals harder, so that her bicycle accelerates up the hill. The question asks students to identify the situations in which the total mechanical energy of the cyclist and bicycle is constant (earth is implicitly part of the system and all students in this study knew about it). On this question, 17% of upper-level students selected all three situations, a percentage similar to that of introductory students (24%). In their written explanations, one upper-level student who provided this response stated, "mechanical energy will be conserved as long as it is not being lost to things like friction and air resistance" and another stated "no other forces used except conservative force like gravity". A second alternate conception that was not as common among upper-level students was selecting only situation 2 (14% of upper-level students), which may be due to only focusing on kinetic energy. For example, one upper-level student who had this type of difficulty motivated their choice by stating that it involves "constant speed".

Q10 describes an explosion in which a bomb at rest on a horizontal surface breaks up into three fragments, all of which fly off horizontally and asks which statements are true among (1) the total kinetic energy of the bomb fragments is the same as that of the bomb before the explosion, (2) the total momentum of the bomb fragments is the same as that of the bomb before explosion, and (3) the total momentum of the bomb fragments is zero. On this question, 29% of upper-level students thought that all three statements are correct. For example, one upper-level student motivated choosing all three by stating "by momentum and energy conservation." It is interesting that students who selected all three recognized that the initial momentum is zero (because the initial velocity is zero), thus could potentially deduce that the initial kinetic energy must be zero as well. The final kinetic energy, however, cannot be zero because all three pieces are moving. However, one upper-level student who selected all three statements reasoned that the kinetic energy stays the same by stating "the amount of total kinetic energy does not change, but just the orientations do. So even if it is represented different the overall [kinetic energy] will be the same." It is possible that some students with this type of response may be considering that kinetic energy has a direction. We also note that on this question, introductory students who answered the question incorrectly selected each incorrect answer choice in comparable percentages (14%, 14%, 18%, and 19%, for answers a, b, c, and e, respectively) whereas most of the upper-level students who answered the question incorrectly, selected answer choice (e) – 29% as mentioned earlier.

Q12 describes a situation in which a constant applied force $F_A$ is used to pull a box with constant velocity along a horizontal surface where there is a force of friction, $F_k$ that is not negligible. On this question, 21% of upper-level students selected (e) which states that "The magnitude of $F_A$ is greater than the magnitude of $F_k$", which is consistent with the thinking that motion at constant velocity requires a net force. For example, one upper-level student stated "since the box moves forward that implies $F_A$ must be bigger than $F_k$ or else it wouldn't move at all".

Q13 relates to the two situations shown in Fig. 7. The carts are identical in all respects and in situation (i), cart A rolls down from an initial height $h$ to the bottom, collides into and sticks to cart B. In situation 2, both carts roll down from a height of $h/2$, collide at the bottom and stick together. Q13 asks which statement is true about the two-cart system just before the two carts collide and provides the following options (a) the kinetic energy of the system is zero in case (ii), (b) the kinetic energy of the system is greater in case (i) than in case (ii), (c) the kinetic energy of the system is the same in both cases, (d) the momentum of the system is greater in case (ii) than in case (i), (e) the momentum of the system is the same in both cases. On this question, 57% of upper-level students selected the correct answer, a percentage nearly identical to that of introductory students (53%). The rest of the upper-level students appeared to be somewhat spread over the four incorrect answers with (b) being the most common, selected by 17% of the students. Explanations suggest that some of the students who selected (b) incorrectly concluded that doubling the height results in doubling the speed at the bottom of the hill (as opposed to doubling the kinetic energy at the bottom of the hill). For example, one upper-level student who provided explanations motivated choosing (b) by stating that "doubling the height of the hill results in a factor of 4 increase in the kinetic energy. Therefore, the cart in system (i) has 4 times as much kinetic energy as either of the individual carts in setup (ii), and still twice as much kinetic energy as the whole system in setup (ii)".

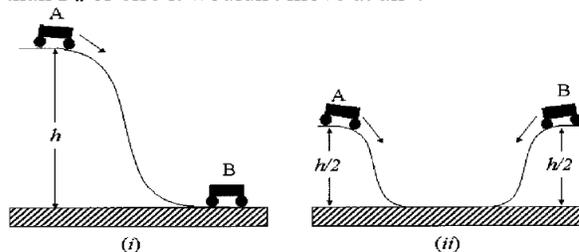

Fig. 7. Diagram for Q13.

In Q16 described earlier (see Fig. 4), 35% of upper-level students selected either only conservation of momentum (21%) or only that the mechanical energy is constant (14%) as the principles needed to solve this problem. Prior research has found that when solving this problem (i.e., given $h_0$, calculate $h_f$), many introductory students have difficulty recognizing that both concepts are necessary and use only one or the other [55], and it appears that many upper-level students had a similar difficulty. Written responses suggest that some students may be answering this question without recognizing the three stages necessary to solve this problem: from when sphere A is released to when it reaches the bottom just before collision, the collision between the spheres, and from just after the collision until the maximum height is reached. When providing explanations for choosing only the conservation of momentum, most upper-level students provided an explanation similar to this student who stated that "Kinetic energy is not conserved in inelastic collisions so it must be [conservation of momentum only]". Some students with this type of reasoning did not seem to recognize that mechanical energy being constant needs to be used prior to the collision and after the collision to relate the heights with the speed of the spheres at the bottom. Furthermore, on this question, introductory students were more likely to select only the mechanical energy being constant as needed to solve the problem compared to only the conservation of momentum (37% compared to 19%). Upper-level students were somewhat more likely to select the conservation of momentum only (21% compared to 14%).

In Q17, a ball is dropped from a high tower and falls freely under the influence of gravity. The question asks students to select the statement that is true among the following: (a) The kinetic energy of the ball increases by equal amounts in equal times, (b) the kinetic energy of the ball increases by equal amounts over equal distances, (c) there is zero work done on the ball by the gravitational force as it falls, (d) the work done on the ball by the gravitational force is negative as it falls, (e) the total mechanical energy of the ball decreases as it falls. On this question, the most common incorrect answers of both groups of students are (a), selected by 27%/17%, and (d), selected by 23%/19% of introductory/upper-level students, respectively. One upper-level student who selected option (a) explained that "since acceleration is constant, the ball gains velocity at a constant rate, thus kinetic energy", which appears to imply that the student did not recognize that the kinetic energy depends on the square of the velocity. Students who selected (d) often reasoned that the force of gravity points in the negative direction or that the direction of motion of the ball is in the negative direction. For example, one upper-level student selected this option explained, "Gravity is a negative force, so the work done by this force will be negative." It also appears that this difficulty regarding associating only the direction of the gravitational force with the sign of the work done by it only exhibited a decrease of 4% in the percentage of students who had this difficulty from the introductory to the upper-level.

In Q22, three balls are launched from the same height at the same speed but at different angles (see Fig. 8). They all reach a dotted line at a certain height above their starting position and students are asked to rank the balls according to their speed when they reach the line. Q22 is the most difficult question on the EMCS: only 29% of upper-level students answered it correctly, which is identical to the percentage of introductory students who answered it correctly. On this question, 33% of upper-level students and 39% of introductory students ranked the speeds (from largest to smallest) in the same order as their $y$ components of velocity. For example, one upper-level student stated:

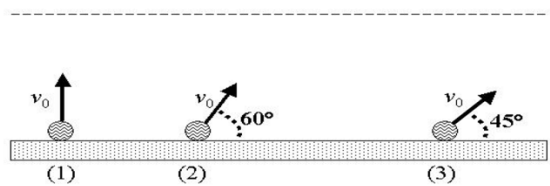

Fig. 8. Diagram for Q22

"The parameter that matters here is the vertical velocity. For [ball 1] it is the biggest, at $v_o$. For [ball 2] it is the second biggest at $v_0 \sin 60 = v_0\sqrt{3}/2$. For [ball 3] it is the smallest at $v_0\sqrt{2}/2$, therefore the order is 1), 2), 3)." Also, written reasonings suggest that sometimes, even if a student recognized that energy considerations could be used, they still appeared to consider that the velocities in the $y$ direction are not the same and incoporated them in their responses. For example, another upper-level student who selected the same order stated: "I imagine the equation KE1=PE+KE2 at the dashed line. If this is correct, then all of the masses cancel out. The fastest ball will be whichever one has the most velocity in the vertical direction." Furthermore, 31% of upper-level students and 16% of introdcutory students ranked them in the same order as their $x$ components of velocity. These students sometimes appear to have focused on the $x$ component of velocity not being affected by the force of gravity, reasoning that the object with the largest initial component of velocity in the $x$ direction will be the one that reaches the dashed line moving fastest. For example, one upper-level student stated: "45 degrees has the highest component of horizontal velocity which gravity cannot affect and so moves more quickly when reaching the horizontal line", and another upper-level student stated: "Vertical speeds are killed by gravity, while horizontal speeds are not." While it is true that ball (1), which is thrown vertically, has more vertical velocity than ball (2), which is thrown at an angle of 60°, we need to consider both components of velocity because the question is asking for the speed. We note that it is possible to use the equations of projectile motion to actually conclude that the total speed when the balls reach the dotted line is the same in all cases despite the different angle because e.g., when comparing balls (1) and (2), the horizontal component of the velocity of ball (2) "compensates" for its lower vertical component. But this is difficult to conclude without doing the actual calculation, whereas if one considers that mechanical energy is constant, it is easier to recognize that there is no difference between the two cases because the directions in which the balls are thrown do not affect their initial kinetic energies. This suggests that a student who does not consider this question from the perspective of energy will have a greater difficulty answering it correctly. It is important to point out just how difficult this question is for upper-level students: 1) it is the most difficult question on the entire survey – only 29% of students answer it correctly in a post-test and 2) it is the only question on the entire survey on which there are *two* incorrect answer choices selected by 30% or more students. Furthermore, the two incorrect options that many students selected relate to considering either that what matters is the horizontal component or the vertical component. This suggests that many students may not be considering the question from the perspective of energy.

Q23 is similar to Q5 in that it also relates to the impulse-momentum theorem: two balls, one made of rubber and the other made of putty, are dropped from the same height. The rubber ball bounces up after the collision with the ground while the putty ball comes to rest. The question also states that the collisions take the same amount of time $\Delta t$ and asks which ball exerts a larger average force on the floor. On this question, 29% of upper-level students stated that the forces are the same. For example, one upper-level student explained: "Average force applied by the surface will equal the change in momentum / dt. It is the same in both cases." Also, many students seemed to answer Q5 and Q23 quite consistently: roughly equal percentages of students answer the questions correctly (52% and 55% for upper-level students on Q5 and Q23, respectively), and the percentages of students with the aforementioned alternate conception is identical as well (29% for upper-level students on both questions).

In Q24, a force $F$ is applied for the same distance $d$ on two blocks of different masses. The question asks which block has a larger kinetic energy. Roughly similar percentages of upper-level students stated that the smaller block will have the larger kinetic energy because of its larger speed (answer choice (b), selected by 14% of students) or that the larger block will have the larger kinetic energy because of its larger mass (answer choice (c), selected by 17% of students). These students did not realize that they can use the work-kinetic energy theorem to reason that the work done by the applied force is the same in the two cases, thus the two objects must have the same kinetic energy. This is somewhat similar to Q22 in which some students focused either on the *x* or the *y* component of velocity rather than considering the bigger picture using energy considerations. In Q24, some students either focused on the mass or on the speed at the expense of the broader picture using energy considerations. Comparing the upper-level students with introductory students on this question, we find that the introductory students also selected answer choice (d) "not enough information" at roughly comparable rates as options (b) and (c) with percentages ranging from 20% to 24%. Upper-level students also primarily selected options (b) and (c) but with smaller frequency, with pertentages ranging from 14% to 17%. Upper-level students who selected option (b), which stated that the kinetic energy is greater for the smaller mass block because it achieves a larger speed, often focused on the fact that the speed is squared when calculating the kinetic energy, and thus the smaller block which has a larger speed would have the larger kinetic energy. This difficulty might be due to the fact that some of these students struggled to distinguish the case of a force being applied for a certain *distance* (suggesting use of the work-kinetic energy theorem) from the case of a force being applied for a certain *time* (suggesting the use of the impulse-momentum theorem). Explanations suggest that some students who answered that the larger block will have the larger kinetic energy appeared to have only focused on the fact that kinetic energy depends on mass and thus the larger block will have a larger kinetic energy, but not on whether the final speeds would be the same. For example, one upper-level student with this response explained their reasoning stating that "kinetic energy is depending on mass".

**SUMMARY, DISCUSSION AND CONCLUSION**

In the research presented here, we investigate how student performance on the EMCS evolves from the beginning to the end of introductory and advanced mechanics courses and then analyze how the performance of the upper-level students compares to the introductory students after traditional instruction. A particular focus was on investigating the extent to which a traditionally taught upper-level undergraduate classical mechanics course, which primarily focused on quantitative problem-solving regarding teaching and assessment, helped students develop conceptual understanding of the introductory concepts covered on the EMCS. We also identified questions on the EMCS which are challenging for both introductory and advanced groups after traditional instruction and also those questions for which there is marked improvement.

The normalized gain from the beginning to the end of the course was 0.3 or larger for eleven questions for introductory students and seven questions for upper-level students. In particular, many upper-level students struggled with introductory classical mechanics concepts on the EMCS after traditional instruction in upper-level classical mechanics. The questions on the EMCS are conceptual in nature and upper-level students had difficulties with several conceptual aspects of energy and momentum covered in the EMCS, even if they were able to carry out complicated calculations in their homework and exam problems, similar to what was found in a study focusing on conceptual understanding of physics graduate students [24]. For example, on the two questions on the EMCS directly related to the impulse momentum theorem, the upper-level students exhibited very low performance of roughly 50%, indicating that they need better scaffolding support in correctly interpreting the impulse momentum theorem and/or recognizing its utility in given situations. Additionally, we find that on all the questions on the post-test which showed a final performance of upper-level students greater than 80%, their pre-test performance was greater than 70%. While it is true that the upper-level students did improve on some questions on which they started very low in pre-test, it was unlikely for them to improve beyond a certain level. Furthermore, on roughly half the questions on the EMCS, the performance of upper-level students either decreased slightly or showed little improvement (less than 5% from the pre-test to post-test). For a few of those questions, this could be explained due to very high pre-test performance, but in most, that was not the case. We also note that on the entire EMCS, the normalized gain of upper-level students was a mere 14%, corresponding to an increase from a 63% average on the pre-test to a 68% average on the post-test.

We find that on all the EMCS questions on which introductory students' performance was less than 50% in the post-test (after instruction), less than two-thirds of the upper-level students provided the correct response in post-test. It appears that on questions in which their incoming knowledge is very low, it is very difficult to improve significantly after traditional instruction in an upper-level course on these concepts. We note that traditional upper-level classical mechanics teaching and assessment did not focus on the kinds of questions that are on the EMCS, although the course deals with the same underlying concepts. Additionally, we find that the upper-level students often displayed the same types of difficulties on the EMCS questions as introductory students, usually with lower percentages, but there wasn't a significant difference between the two groups on some of the questions. These findings suggest that to help students develop a functional understanding, upper-level classical mechanics courses should use research-based approaches that integrate conceptual and quantitative aspects of mechanics similar to introductory courses. As Mazur noted [56, 57], students can become very adept at regurgitating solution patterns using memorized algorithms, but not be able to answer 'simpler' questions, e.g., comparing the brightness of different light bulbs in significantly simpler circuits. If instruction does not explicitly integrate both conceptual and quantitative problem solving in teaching and assessment, students can rely on algorithmic problem-solving approaches and perform well despite lacking functional understanding of the underlying physics concepts.

In summary, consistent with prior studies, these findings highlight that traditional upper-level classical mechanics courses are not effective in helping all students develop a functional understanding of the energy and momentum concepts covered in the EMCS. Instructors should not take for granted that students at the upper-level will make the conceptual and quantitative connections on their own if there is no intentional focus on this type of integration in their teaching and assessment [58, 59]. Instructors should focus on integrating both quantitative and conceptual aspects of classical mechanics in teaching and assessment using research-based approaches even in upper-level courses. We note that we previously conducted interviews with physics instructors who had taught traditional upper-level undergraduate and graduate core courses. We found that some instructors incorrectly believe that learning physics concepts is easier for students at all levels than learning how to solve physics problems using "rigorous" mathematics [24]. They believe that if non-science majors can learn physics concepts, science, and engineering majors and particularly physics majors can learn physics concepts on their own even if there was no conceptual assessment in the course and there is not much use in instructors wasting precious instructional time explicitly on concepts in advanced courses. Instructors with these types of beliefs often noted that they focus mainly on quantitative problem solving that will help students do complex calculations. Moreover, interviews suggest that in upper-level or graduate courses, many instructors believe that students should have learned the concepts in the previous physics courses (e.g., in introductory courses for upper-level courses or in undergraduate courses for graduate-level courses) so their goal as instructors is mainly to focus on developing the "calculational" facility of students in the courses they are teaching instead of striving to appropriately integrate conceptual and quantitative understanding [24]. Some instructors also believe that students will focus on the physics concepts involved anyway in order to be able to do the calculations meaningfully so there is no need to reward them for conceptual understanding by asking them conceptual questions in assessment tasks [24]. Other instructors claimed that they always mention important concepts involved before doing calculational problems or before doing complicated derivations in their classes [24]. However, they only ask students to do calculations in assessment tasks that determine their course grade (these are often problems that many students learn to do without deep understanding of the basic underlying concepts). Our findings here in the context of EMCS suggest that many upper-level students did not develop a robust understanding of underlying concepts at the end of the course.

Also, even if students have had one round of exposure to concepts in previous courses, explicitly integrating conceptual and quantitative aspects of physics in instructional goals, instructional design and assessment of learning is critical for many students to be motivated to focus on developing functional understanding. Students need to be supported to develop a functional understanding of physics by solving a variety of integrated conceptual-quantitative problems that are appropriately scaffolded. Some instructors in our earlier interviews also claimed that they select calculational problems that have rich conceptual implications although they expect students to unpack those conceptual implications on their own when doing the calculation instead of explicitly integrating conceptual questions with those calculational problems to provide scaffolding support to make appropriate math-physics connections. Without such incentive and support, many students at all levels are not able to make such connections automatically and development of functional understanding is compromised. Promisingly, instructors sometimes noted that they would be willing to incorporate good conceptual questions or integrated conceptual-quantitative problems in their instruction, should such questions be available for them, but that they do not have the time to devote to creating these types of questions [24].

We hope that instructors of traditionally taught upper-level classical mechanics courses will use the findings of the research presented here as motivation to better integrate conceptual and quantitative aspects of classical mechanics in teaching and assessment. We also hope that physics education researchers will develop integrated conceptual-quantitative questions for upper-level classical mechanics courses that are cross-linked to commonly used textbooks to make it easy for the upper-level instructors to adapt such questions in their instructional design and assessment.

## ACKNOWLEDGMENTS

We thank all students whose data were analyzed and Dr. Robert P. Devaty for his feedback on the manuscript.

## APPENDIX: INTRODUCTORY AND UPPER-LEVEL STUDENTS' PERFORMANCE ON PRETEST

**Table III.** Percentages of introductory students and upper-level students who selected each answer choice on each question on the EMCS before instruction (pre-test). The correct answers are bold, underlined, and italicized.

| EMCS Q# | Introductory Students (Pre-test) | | | | | Upper-Level Students (Pre-test) | | | | |
|---|---|---|---|---|---|---|---|---|---|---|
| | A | B | C | D | E | A | B | C | D | E |
| 1 | 10 | ***33*** | 37 | 9 | 11 | 0 | ***76*** | 18 | 4 | 1 |
| 2 | 15 | 7 | 21 | 7 | ***50*** | 15 | 1 | 3 | 4 | ***76*** |
| 3 | 13 | ***45*** | 16 | 16 | 10 | 1 | ***65*** | 0 | 22 | 12 |
| 4 | ***47*** | 14 | 3 | 15 | 21 | ***71*** | 3 | 4 | 7 | 15 |
| 5 | 15 | 45 | 17 | ***11*** | 12 | 9 | 31 | 7 | ***47*** | 6 |
| 6 | 5 | 4 | ***12*** | 66 | 13 | 0 | 4 | ***47*** | 25 | 24 |
| 7 | 15 | 5 | 10 | 15 | ***55*** | 1 | 1 | 4 | 6 | ***87*** |
| 8 | 6 | 8 | ***30*** | 41 | 15 | 6 | 9 | ***57*** | 21 | 7 |
| 9 | ***26*** | 33 | 14 | 12 | 15 | ***49*** | 18 | 10 | 0 | 24 |
| 10 | 35 | 20 | 21 | ***13*** | 11 | 1 | 7 | 16 | ***57*** | 18 |
| 11 | 13 | 23 | 12 | 11 | ***41*** | 1 | 3 | 6 | 4 | ***85*** |
| 12 | 5 | 4 | 9 | ***21*** | 61 | 6 | 1 | 21 | ***47*** | 25 |
| 13 | 6 | 18 | ***52*** | 10 | 14 | 4 | 26 | ***60*** | 6 | 3 |
| 14 | 6 | 14 | 8 | ***60*** | 12 | 0 | 3 | 3 | ***90*** | 4 |
| 15 | ***47*** | 6 | 39 | 4 | 4 | ***69*** | 3 | 26 | 1 | 0 |
| 16 | 24 | 22 | ***34*** | 12 | 8 | 35 | 13 | ***44*** | 1 | 6 |
| 17 | 31 | ***30*** | 5 | 28 | 6 | 21 | ***54*** | 1 | 24 | 0 |
| 18 | 23 | 17 | 19 | 21 | ***20*** | 1 | 4 | 6 | 6 | ***82*** |
| 19 | 14 | ***47*** | 7 | 14 | 18 | 19 | ***72*** | 1 | 6 | 1 |
| 20 | ***35*** | 11 | 33 | 8 | 13 | ***78*** | 0 | 10 | 3 | 9 |
| 21 | 13 | 15 | ***44*** | 11 | 17 | 18 | 1 | ***71*** | 3 | 7 |
| 22 | 35 | 7 | 21 | ***16*** | 21 | 25 | 1 | 34 | ***32*** | 7 |
| 23 | 45 | ***21*** | 20 | 9 | 5 | 34 | ***51*** | 13 | 1 | 0 |
| 24 | ***20*** | 29 | 29 | 15 | 7 | ***50*** | 19 | 19 | 10 | 1 |
| 25 | 15 | 13 | 19 | 20 | ***33*** | 7 | 7 | 12 | 10 | ***63*** |